\documentclass{article}
\usepackage{spconf,amsmath,graphicx,hyperref,booktabs,multirow}
\usepackage[table]{xcolor}

\newcommand{\huggingfacesmall}{\includegraphics[width=9px]{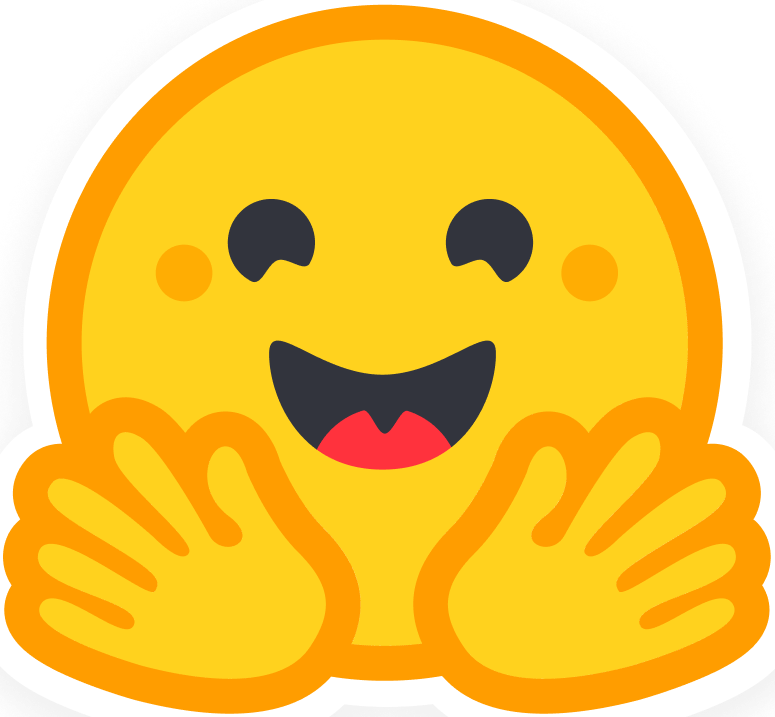}}
\newcommand{\githubsmall}{\includegraphics[width=9px]{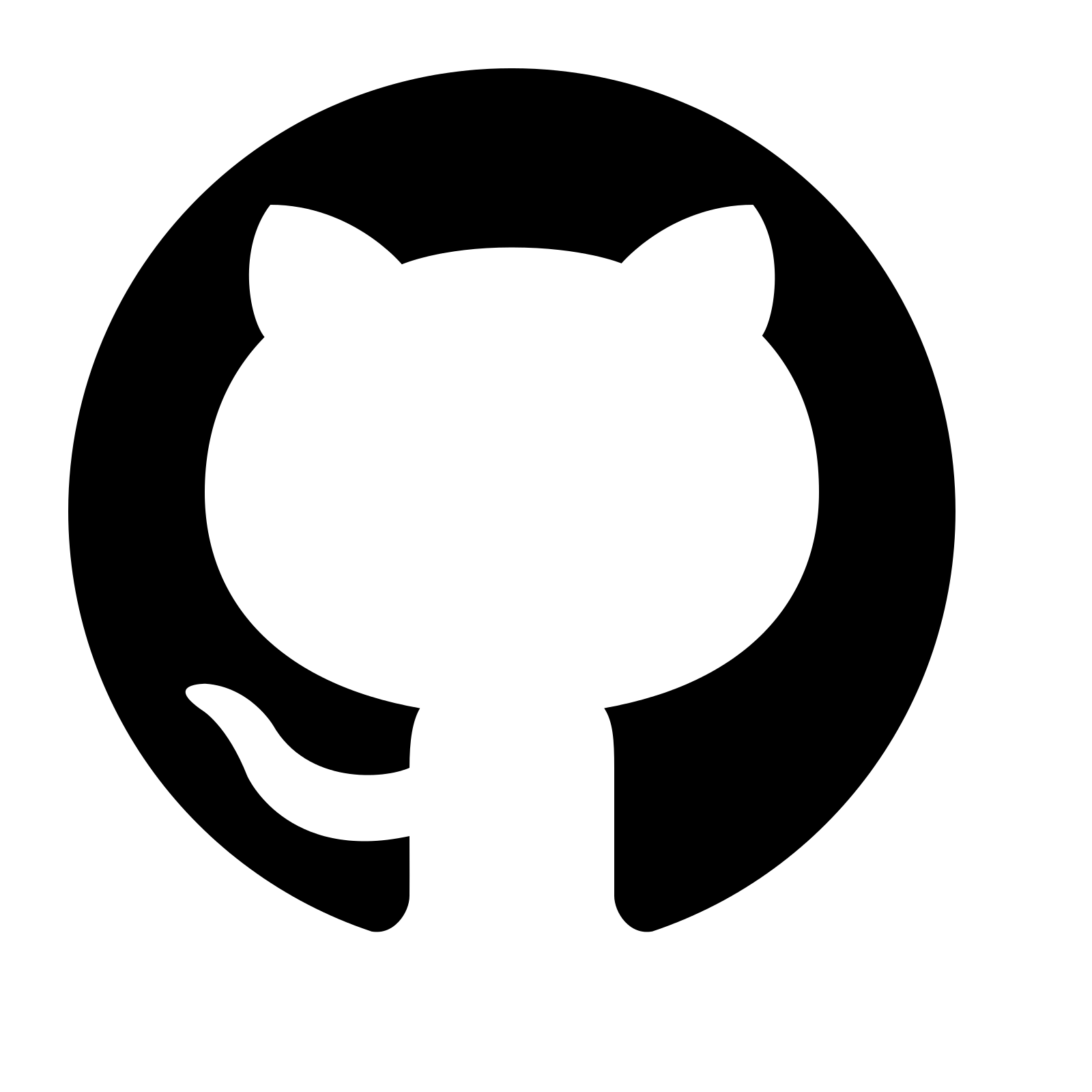}}
\title{LOUD AND CLEAR: DYNAMIC ACTIVATION STEERING FOR IMPROVING SPEECH INTELLIGIBILITY IN NOISY ENVIRONMENTS}
\name{Seymanur Akti$^{1,2}$, Alexander Waibel$^{1,3}$}
\address{$^{1}$Karlsruhe Institute of Technology (KIT), Karlsruhe, Germany\\
$^{2}$KIT Campus Transfer (KCT), Karlsruhe, Germany\\
$^{3}$Carnegie Mellon University (CMU), Pittsburgh, USA}
\begin{document}
%
\maketitle
\begin{abstract}
Speech becomes less intelligible in noisy environments, and humans naturally adapt their voice to compensate. Inspired by this behavior, we investigate whether a text-to-speech (TTS) model can be guided to produce more intelligible speech using activation steering, without retraining. We focus on two characteristics of the Lombard effect: increased vocal effort and hyper-articulation. We introduce a prompt-relative steering mechanism that prevents steering effects from accumulating during generation while allowing their strength to be adjusted dynamically. Across seen and unseen speakers and multiple languages, our method produces systematic changes in Lombard-related acoustic features, preserves speaker similarity (89--95\%), and reduces WER under background noise by 7--22\% at 1 dB SNR. These results show that pretrained TTS models can be dynamically controlled to generate more intelligible speech without retraining.

\end{abstract}

\section{Introduction}
\label{sec:intro}

Humans naturally adapt their speaking style to challenging acoustic conditions such as noise, a phenomenon known as the Lombard effect. These adaptations improve speech intelligibility through increased vocal effort, higher pitch, slower speaking rate, flattened spectral tilt, and clearer articulation~\cite{junqua1993lombard, soltau2000specialized}. Such articulatory changes have also been shown to be relevant to robust speech recognition~\cite{metze2002flexible}. Reproducing such adaptations in synthesized speech could benefit speech synthesis in noisy environments and applications requiring enhanced intelligibility, such as second-language (L2) learning and assistive communication.

Early near-end listening enhancement (NELE) and Lombard synthesis relied on signal-processing techniques, including spectral modification~\cite{brouckxon2008time,sauert2012near,taal2013sii,taal2014speech}, dynamic-range compression~\cite{zorila2014spectral,kandia2012speech}, and prosodic manipulation~\cite{raitio2011analysis}, often introducing acoustic distortions. Recent neural TTS approaches instead use transfer learning~\cite{bollepalli2019lombard,paul2020enhancing}, fine-tuning~\cite{hu2021whispered}, spectral-tilt augmentation~\cite{raitio2022vocal}, explicit pitch/energy modeling~\cite{woszczyk2025voice}, ASR-guided loudness control~\cite{novitasari2021dynamically}, and latent-space interpolation~\cite{lobato2025gradual}. These methods primarily target global acoustic properties such as vocal effort, pitch, and spectral characteristics, while hyper-articulation remains less explored. Articulation has been modeled using HMMs~\cite{picart2020analysis} and Global Style Tokens (GSTs)~\cite{nishihara2024low}, but vocal effort and hyper-articulation have generally been treated as separate control dimensions. A recent study proposed joint continuous control of both attributes at the utterance and word levels~\cite{akti2026synthesizing}.

In real-world settings, the required degree of intelligibility enhancement may vary with acoustic conditions or listener requirements. Human speakers continuously adapt their speaking style to environmental feedback, motivating TTS systems that can dynamically adjust intelligibility-related attributes, potentially within an utterance. Prior work has explored incremental speech transformation~\cite{rottschafer2015online} and ASR-based feedback for dynamic loudness control~\cite{novitasari2021dynamically}, but does not provide joint, continuous, within-utterance control of vocal effort and hyper-articulation. Moreover, existing Lombard TTS methods generally rely on training, fine-tuning, or fixed signal-level transformations, limiting dynamic inference-time control of multiple attributes across speakers and languages.

We address these limitations with inference-time activation steering, which modifies hidden representations along attribute-specific directions without changing model parameters~\cite{turner2023steering}. Recent work has applied activation steering to TTS for emotion control~\cite{zhou2026emoshift,wang2026cocoemo,xie2025emosteer} and accent neutralization~\cite{yang2026activation}, but repeatedly applying a fixed intervention during autoregressive generation can accumulate across tokens and distort the output~\cite{kang2026prompt}. To overcome this limitation, we introduce \textit{Prompt-Relative Activation Steering}, which compares each generated token's activation with its baseline activation from the reference prompt and applies only the residual steering needed to reach the target. This prevents steering accumulation while enabling stable, continuous control. Using paired speech data, we extract activation directions for vocal effort and hyper-articulation from Qwen3-TTS~\cite{hu2026qwen3} and evaluate the approach across multiple speakers and languages. Audio samples are shared at demo page.~\footnote{https://seymanurakti.github.io/loud-and-clear/}

Our contributions are threefold: (1) joint and continuous control of vocal effort and hyper-articulation through training-free activation steering, demonstrated across speakers and languages, (2) a prompt-relative mechanism that prevents steering accumulation during autoregressive generation, and (3) dynamic within-utterance control, allowing steering strength to be modified during synthesis.


\vspace{-0.2cm}
\section{Method}
\label{sec}
\vspace{-0.2cm}
Following prior work on activation steering~\cite{xie2025emosteer,wang2026cocoemo,yang2026activation}, we extract attribute-specific steering directions from paired speech conditions and apply them to hidden representations during inference. We use Qwen3-TTS and intervene on the residual outputs of its LLM-based autoregressive backbone. We consider two intelligibility-related attributes: vocal effort and hyper-articulation. Their steering directions are extracted independently and combined for joint control. The control is applied only to generated tokens to enable dynamic control in streaming mode.
\vspace{-0.2cm}

\subsection{Steering Vector Extraction}
\label{sec}

We first synthesize paired speech samples using the voice-cloning generation method of Qwen3-TTS. Each pair is produced by the same speaker with identical linguistic content but different speaking styles, isolating the target attribute.

Let $\mathbf{h}_{l,t}^{(x,i)}$ denote the hidden representation at layer $l$ and generated-token position $t$ for utterance $i$ with speaking style $x$. We obtain an utterance-level representation by averaging over generated tokens:
\vspace{-0.3cm}

\begin{equation}
\mathbf{h}_{l,i}^{(x)}=\frac{1}{T_i}
\sum_{t=1}^{T_i}
\mathbf{h}_{l,t}^{(x)}
\end{equation}
\vspace{-0.2cm}

where $T_i$ is the number of generated tokens for utterance $i$. The steering direction for attribute $k$ at layer $l$ is computed as the average difference between the target and reference speaking styles across $N_k$ paired samples:
\vspace{-0.1cm}
\begin{equation}
\mathbf{v}_{l}^{k} = \frac{1}{N_k}\sum_{i=1}^{N_k} (\mathbf{h}_{l,i}^{(target)} - \mathbf{h}_{l,i}^{(reference)})
\end{equation}
where $k\in\{E,A\}$ denotes vocal effort and hyper-articulation, respectively. For vocal effort, the target and reference conditions correspond to loud and regular speech from EARS dataset~\cite{richter2024ears}, while for hyper-articulation they correspond to enunciated and default speech from Expresso dataset~\cite{nguyen2023expresso}. The directions are extracted only from generated-token positions rather than input prompt tokens, so that they characterize the representations used during speech generation and can be applied directly to the autoregressive generation process.
\vspace{-0.2cm}
\subsection{Joint Control of Vocal Effort and Articulation}
\label{sec:joint_steering}
The two attributes are controlled jointly by combining their layer-wise steering directions. Let $\mathbf{v}_l^{E}$ and $\mathbf{v}_l^{A}$ denote the vocal-effort and articulation directions, respectively and $\lambda_E$ and $\lambda_A$ are the user-defined control coefficients:

\vspace{-0.1cm}
\begin{equation} 
\mathbf{v}_l^{\mathrm{joint}}= \lambda_E \mathbf{v}_l^{E} + \lambda_A \mathbf{v}_l^{A} 
\end{equation} 

\vspace{-0.1cm}
This formulation provides continuous control over the relative contribution of vocal effort and articulation. In particular, setting one coefficient to zero produces single-attribute steering, while non-zero values for both coefficients provide joint continuous control.

\vspace{-0.2cm}
\subsection{Prompt-Relative Activation Steering}
\label{sec:autoregressive_steering}

Adding a fixed steering vector to every generated token can lead to cumulative changes in the hidden representations, as each token already inherits part of the steering effect through causal attention. Consequently, the generated audio gets distorted as the sequence becomes longer. To mitigate this effect, we perform prompt-relative steering.

During prompt processing, we compute the mean projection of the hidden representations corresponding to the reference audio segment onto the normalized joint steering direction. This projection  $p_{l}^{baseline}$ serves as the prompt-specific baseline for subsequent generation. The target projection $p^{*}_{l}$ is then defined as the baseline projection plus the magnitude of the weighted joint steering vector at each layer.
\vspace{-0.2cm}
\begin{equation}
p_{l}^{baseline}=
\frac{1}{T_p}
\sum_{t=1}^{T_p}
\mathbf{h}_{l,t}^{\top}
\frac{\mathbf{v}_l^{\mathrm{joint}}}{||\mathbf{v}_l^{\mathrm{joint}}||_2}
\end{equation}

\begin{equation}
p^{*}_{l}=p_l^{baseline} + ||\mathbf{v}_l^{\mathrm{joint}}||_2
\end{equation}

During generation, for each hidden representation $\mathbf{h}_{l,t}$, we compute its projection onto the unit joint steering vector, representing its current alignment with the steering direction.

\begin{equation}
p_{l,t}=
\mathbf{h}_{l,t}^{\top}
\frac{\mathbf{v}_l^{\mathrm{joint}}}{||\mathbf{v}_l^{\mathrm{joint}}||_2}
\end{equation}

The required steering magnitude is calculated as the difference between the target and current projections, and the residual is applied along the unit joint steering direction:

\begin{equation}
\tilde{\mathbf{h}}_{l,t}=
\mathbf{h}_{l,t}
+
(p^{*}_{l}-p_{l,t})\frac{\mathbf{v}_l^{\mathrm{joint}}}{||\mathbf{v}_l^{\mathrm{joint}}||_2}
\end{equation}

The residual steering is signed and can therefore be either steering the token in positive or negative direction. This is important for online control: when the steering strengths are reduced during generation, the target projection is reduced accordingly, and a negative residual removes part of the previously applied steering, allowing the hidden representation to move back toward the prompt baseline.

Finally, the modified representation is normalized to preserve the original hidden-state magnitude. The steering procedure is applied to selected Transformer layers (19--20) during autoregressive inference.

\vspace{-0.2cm}

\section{Experiments and Results}

The activation steering vectors were extracted from the regular and loud subsets of EARS~\cite{richter2024ears} (107 speakers) and the default and enunciated subsets of Expresso~\cite{nguyen2023expresso} (4 speakers). We follow the same evaluation protocol with SLE~\cite{akti2026synthesizing}, synthesizing first 5 lists of Harvard sentences~\cite{rothauser1969ieee} for four Expresso speakers. We additionally evaluate generalization to unseen speakers using the first 10 speakers of VCTK and to unseen languages using the German, Spanish, and Japanese test sets from TTS Multilingual~\footnote{\huggingfacesmall \href{https://huggingface.co/datasets/MiniMaxAI/TTS-Multilingual-Test-Set}{MiniMaxAI/TTS-Multilingual-Test-Set}}.

We use acoustic and intelligibility measures to assess the effects of steering. \textbf{Vowel Space Area (VSA)} is the area spanned by the corner vowels /a/, /i/, and /u/ in the F1--F2 space, correlated with increased articulatory distinctiveness. Phoneme alignments were obtained by the Montreal Forced Aligner~\footnote{\githubsmall \href{https://github.com/MontrealCorpusTools/Montreal-Forced-Aligner}{MontrealCorpusTools/Montreal-Forced-Aligner}}, with formants extracted using Parselmouth~\footnote{\githubsmall \href{https://github.com/YannickJadoul/Parselmouth}{YannickJadoul/Parselmouth}}. \textbf{Spectral Tilt (ST)} is computed as the ratio of spectral energy in the 1--5 kHz band to that below 1 kHz and reflects changes associated with vocal effort. \textbf{Phoneme Rate (PR)} is the number of phonemes produced per second. \textbf{Speaker Similarity (SSIM)} is measured as the cosine similarity between WavLM speaker embeddings~\footnote{\huggingfacesmall \href{https://huggingface.co/microsoft/wavlm-base-sv}{microsoft/wavlm-base-sv}} of synthesized and reference speech.\textbf{Word Error Rate (WER)} is computed using Whisper-medium~\footnote{\githubsmall \href{https://github.com/openai/whisper}{openai/whisper}} as an intelligibility measure.
\vspace{-0.3cm}

\begin{table}[h]
\centering
\small
\setlength{\tabcolsep}{4.5pt}
\caption{Performance on the Expresso dataset. Highlighted rows indicate proposed configurations.}
\label{tab:expresso_results}
\begin{tabular}{lccccc}
\toprule
\textbf{Condition} & \textbf{WER $\downarrow$} & \textbf{ST $\uparrow$} & \textbf{VSA $\uparrow$} & \textbf{PR $\downarrow$} & \textbf{SSIM $\uparrow$} \\
\midrule
\rowcolor{gray!12}\multicolumn{6}{l}{\textit{\textbf{SLE} \cite{akti2026synthesizing}}} \\
Baseline & 4.86 & -19.89 & 5.51 & 15.97 & 0.86 \\
Half scaling & 0.70 & -17.81 & 7.71 & 12.36 & 0.86 \\
Full scaling & 1.09 & -14.70 & 7.30 & 10.63 & 0.84 \\
\midrule
\rowcolor{gray!12}\multicolumn{6}{l}{\textit{\textbf{Ours}}} \\
Baseline & 1.89 & -18.87 & 4.18 & 15.14 & 0.90 \\
Low ($\lambda_{E,A} = 0.5$) & 1.77 & -16.61 & 5.35 & 14.13 & 0.90 \\
Mid ($\lambda_{E,A} = 1.0$) & 1.41 & -15.89 & 5.77 & 13.25 & 0.90 \\
High ($\lambda_{E,A} = 2.0$) & 1.22 & -15.03 & 5.84 & 12.30 & 0.89 \\
\midrule
Vocal effort only & 1.81 & -14.76 & 5.52 & 15.06 & 0.90 \\
Articulation only & 1.71 & -18.83 & 5.28 & 12.36 & 0.89 \\
\bottomrule
\end{tabular}
\vspace{-0.3cm}
\end{table}

\begin{table}[h]
\centering
\small
\setlength{\tabcolsep}{3.5pt}
\caption{Multilingual and multispeaker evaluation across VCTK, German, Spanish, and Japanese test sets.}
\label{tab:multilingual_results}
\begin{tabular}{llccccc}
\toprule
\textbf{Test Set} & \textbf{Condition} & \textbf{WER $\downarrow$} & \textbf{ST $\uparrow$} & \textbf{VSA $\uparrow$} & \textbf{PR $\downarrow$} & \textbf{SSIM $\uparrow$} \\
\midrule
\multirow{2}{*}{VCTK} 
& Baseline & 0.51 & -21.19 & 1.31 & 16.88 & 0.96 \\
&\cellcolor{gray!12}Steered & \cellcolor{gray!12}0.29 &\cellcolor{gray!12}-17.41 &\cellcolor{gray!12}1.93 &\cellcolor{gray!12} 14.93 &\cellcolor{gray!12}0.95 \\
\midrule
\multirow{2}{*}{German} 
& Baseline & 0.77 & -18.18 & 9.05 & 14.22 & 0.96 \\
&\cellcolor{gray!12}Steered & \cellcolor{gray!12}1.26 &  \cellcolor{gray!12}-13.04 &  \cellcolor{gray!12}4.00 &  \cellcolor{gray!12}11.59 & \cellcolor{gray!12}0.95 \\
\midrule
\multirow{2}{*}{Spanish} 
& Baseline & 1.76 & -22.91 & 5.09 & 16.88 & 0.98 \\
&\cellcolor{gray!12}Steered &\cellcolor{gray!12}1.80 &\cellcolor{gray!12}-18.29 &\cellcolor{gray!12}6.76 &\cellcolor{gray!12}10.93 &\cellcolor{gray!12}0.95 \\
\midrule
\multirow{2}{*}{Japanese} 
& Baseline & 4.68 & -19.61 & 2.13 & 12.35 & 0.96 \\
&\cellcolor{gray!12}Steered &\cellcolor{gray!12}5.64 &\cellcolor{gray!12}-15.88 &\cellcolor{gray!12}2.32 &\cellcolor{gray!12}11.95 &\cellcolor{gray!12}0.94 \\
\bottomrule
\end{tabular}
\vspace{-0.3cm}
\end{table}

\vspace{-0.2cm}
\subsection{Lombard-Related Acoustic Changes}

We evaluate multiple steering strengths, with $\lambda_E=\lambda_A=2.0$ being the maximum safe strength observed experimentally, thereby also testing extrapolation beyond the reference data. The baseline is unsteered Qwen3-TTS speech.

Table~\ref{tab:expresso_results} shows systematic changes with increasing steering strength: spectral tilt becomes less negative, speaking rate decreases, and vowel-space dispersion increases. These trends are consistent with increased vocal effort and hyper-articulation. The individual directions produce distinct effects, with vocal-effort steering primarily affecting spectral tilt and articulation steering producing a stronger reduction in speaking rate. Both directions also increase VSA, while joint steering combines these effects. Compared to SLE~\cite{akti2026synthesizing}, Qwen3-TTS baseline achieves lower WER ($\Delta=-2.97$) for regular speech and maintains slightly higher speaker similarity across steering conditions. The proposed method produces comparable changes in spectral tilt, speaking rate, and VSA relative to their corresponding baselines. While SLE was trained with the same speakers, our model achieves comparable results without additional training, suggesting potential generalization to other speakers.

We further evaluate generalization to unseen VCTK speakers and to German, Spanish, and Japanese, as results shown in Table~\ref{tab:multilingual_results}. The steering directions consistently modify spectral tilt and speaking rate across speakers and languages, although VSA changes are more speaker- and language-dependent. This is expected since the acoustic realization of hyper-articulation depends on phonological structure. The multilingual results nevertheless provide evidence of cross-lingual transfer, while suggesting that precise articulatory control may benefit from language-specific calibration.

Speaker similarity remains largely preserved across steering conditions, indicating that the method can modify Lombard-related attributes without compromising speaker identity.

\begin{figure*}[h]

    \includegraphics[width=\textwidth]{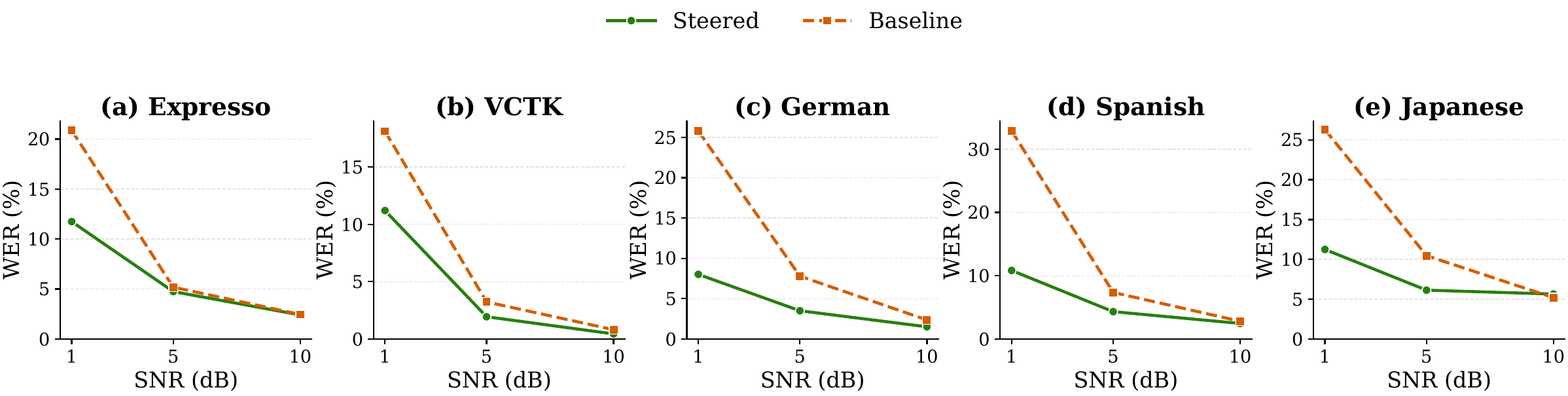}
    \caption{WER results in background noise with different SNR levels.}
    \label{fig:wer_results}
    \vspace{-0.45cm}
\end{figure*}

\vspace{-0.2cm}
\subsection{Speech Intelligibility In Noise}
We evaluate speech intelligibility under restaurant babble noise at SNRs of 1, 5, and 10 dB using WER. For each condition, the noise is scaled to the speech power to maintain a fixed SNR across conditions and avoid attributing intelligibility differences to overall energy gain which could be simply achieved by turning on the device volume. The results in Figure~\ref{fig:wer_results} show that WER consistently decreases when steering is applied. The effect is more prominent in higher noise levels, especially for German, Spanish and Japanese. 

\vspace{-0.2cm}
\subsection{Human Evaluation}

We conduct a user study to evaluate the perceived intelligibility of the proposed method. Participants evaluate 15 pairs of baseline and steered utterances presented with background noise at an SNR of 5~dB. For each pair, participants rate their preference on a scale from $-3$ to $+3$, where $+3$ indicates a strong preference for the steered audio and $-3$ indicates a strong preference for the baseline. Participants are not informed which sample corresponds to which system, and the order is randomized. Seven participants completed the evaluation, resulting in a Comparison Mean Opinion Score (CMOS) of \textbf{+0.958 $\pm$ 0.48} with a 95\% confidence interval. The positive CMOS indicates a preference for the proposed method and provides subjective evidence consistent with the objective intelligibility results.

\vspace{-0.3cm}
\subsection{Dynamic Steering Evaluation}
We adapt the steering method to Qwen3-TTS streaming mode to simulate dynamic steering-strength adaptation. We synthesize 10 long sentences from LibriTTS~\cite{koizumi2023libritts} (12--18 seconds) using a single Expresso reference speaker. Steering is applied only during an interval of the utterance, after which the steering strength returns to zero. We measure spectral tilt changes over time and the additional latency introduced by token-level steering during streaming generation.

\vspace{-0.4cm}
\begin{table}[h]
\centering
\small
\setlength{\tabcolsep}{8pt}
\caption{Streaming performance across evaluation modes.}
\label{tab:latency_results}
\begin{tabular}{lcc}
\toprule
\textbf{Steering Mode} & \textbf{Mean TTFA (s)} & \textbf{Mean RTF} \\
\midrule
No Steering    & 0.469 & 1.078 \\
Dynamic Steering & 0.474 & 1.088 \\
Constant Steering            & 0.475 & 1.086 \\
\bottomrule
\end{tabular}
\vspace{-0.2cm}
\end{table}

As shown in Table~\ref{tab:latency_results}, our streaming activation steering mechanism maintains real-time efficiency, confirming that the streaming hooks add negligible computational overhead. Meanwhile, Table~\ref{tab:intermediate_spectral_tilt} demonstrates temporal precision: dynamic steering successfully follows the no steering baseline prior to intervention, shifts instantaneously during the active steering window and matches constant steering behavior, and cleanly reverts post-intervention.

\begin{table}[h]
  \centering
  \caption{Segmented spectral tilt (dB) across steering modes.}
  \label{tab:intermediate_spectral_tilt}
  \small
  \begin{tabular}{lccc}
    \toprule
    \textbf{Steering Mode} & \textbf{Before} & \textbf{During} & \textbf{After} \\
    \midrule
    No Steering         & -20.84 & -21.06 & -20.96 \\
    Dynamic Steering    & -20.51 & \textbf{-16.94} & -20.68 \\
    Constant Steering   & -17.75 & -17.26 & -17.65 \\
    \bottomrule
  \end{tabular}
  \vspace{-0.4cm}
\end{table}

\begin{figure}[h]
    \centering
    \includegraphics[width=\linewidth]{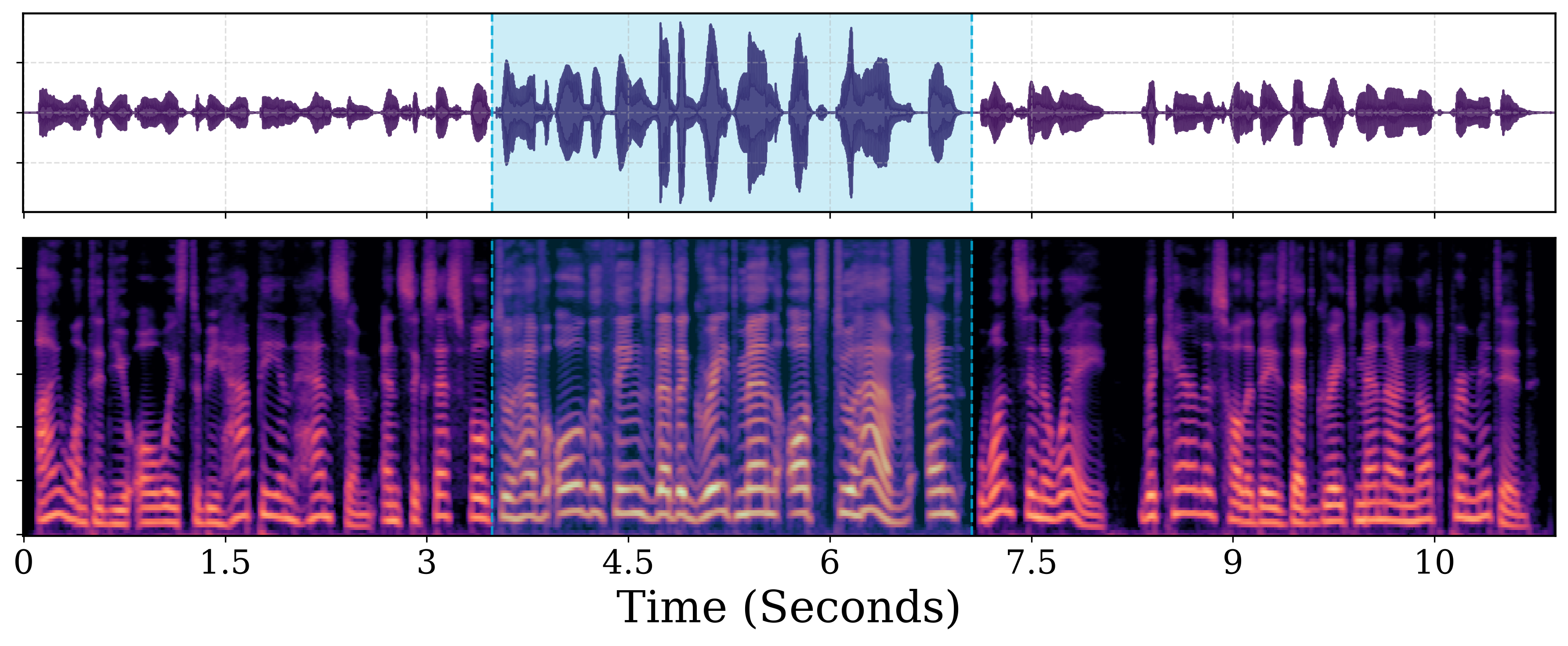}
    \caption{Features of the dynamically steered speech. Highlighted part corresponds to active steering window.}
    
    \label{fig:intermediate_steering_mel}
    \vspace{-0.2cm}
\end{figure}

Figure~\ref{fig:intermediate_steering_mel} shows an example of dynamically steered speech. The increase in amplitude and shift in energy toward higher frequencies during steering correspond to the simulated Lombard effect. After the steering interval ends, speech returns to its baseline amplitude and frequency distribution, demonstrating dynamic control during streaming.

\vspace{-0.2cm}

\vspace{-0.4cm}
\section{Conclusion}
\vspace{-0.1cm}
We presented a training-free activation steering approach for controllable Lombard speech synthesis with Qwen3-TTS. Using paired speech data, our method jointly controls vocal effort and hyper-articulation without model fine-tuning, with experiments across seen and unseen speakers and multiple languages. To prevent cumulative effects in autoregressive generation, we introduced \textit{Prompt-Relative Activation Steering}, which regulates each activation relative to the reference voice and enables stable, dynamic control within an utterance. Experiments demonstrate systematic Lombard-related acoustic changes, preserved speaker similarity, and improved ASR-based intelligibility under noise. Future work will address more diverse acoustic environments and listener types and improve hyper-articulation control across languages.

\let\oldbibliography\thebibliography
\renewcommand{\thebibliography}[1]{%
  \oldbibliography{#1}%
  \small
  \setlength{\itemsep}{2pt}%
  \setlength{\parskip}{2pt}%
}
\bibliographystyle{IEEEtran}
\bibliography{refs}

\end{document}